\documentclass[%
 reprint,
superscriptaddress,
showpacs,preprintnumbers,
 amsmath,amssymb,
 aps,
]{revtex4-1}

\usepackage{graphicx}
\usepackage{dcolumn}
\usepackage{bm}

\usepackage[caption=false]{subfig}
\usepackage{epstopdf}

\newcommand{\bra}[1]{\left\langle{#1}\right|}
\newcommand{\ket}[1]{\left|{#1}\right\rangle}

\newcommand\redout{\bgroup\markoverwith{\textcolor{red}{\rule[.5ex]{2pt}{0.4pt}}}\ULon}

\usepackage[colorlinks=true,citecolor=blue]{hyperref}
\hypersetup{colorlinks=true,citecolor=blue,linkcolor=red,urlcolor=blue}

\begin{document}
\title{Magnetotransport of a 2DEG with anisotropic Rashba interaction at the LaAlO$_3$/SrTiO$_3$ interface }
\author{Azadeh Faridi}
\affiliation{Department of Physics, Sharif University of Technology, Tehran, Iran}
\affiliation{School of Physics, Institute for Research in Fundamental Sciences (IPM), Tehran 19395-5531, Iran}

\author{Reza Asgari}
\email{asgari@ipm.ir}
\affiliation{School of Physics, Institute for Research in Fundamental Sciences (IPM), Tehran 19395-5531, Iran}
\affiliation{School of Nano Science, Institute for Research in Fundamental Sciences (IPM), Tehran 19395-5531, Iran}

\author{Abdollah Langari}
\affiliation{Department of Physics, Sharif University of Technology, Tehran, Iran}
\affiliation{School of Physics, Institute for Research in Fundamental Sciences (IPM), Tehran 19395-5531, Iran}

\date{\today}

\begin{abstract}
We investigate the magnetotransport properties of a two-dimensional electron gas with anisotropic k-cubic Rashba interaction at the $\rm{LaAlO_3}$/$\rm{SrTiO_3}$ interface. The Landau levels and density of states of the system as well as the magnetotransport coefficients are evaluated. A somehow anomalous beating pattern in low magnetic field regime is found both in the density profile and magnetoresistivity. We discuss the impact of electron density, Landau level broadening and Rashba spin-orbit constant on the appearance of the beatings in low magnetic fields and find that at low electron concentrations and not very strong spin-orbit interactions the beatings smooth out. On the other hand, as the magnetic field increases, the Zeeman term becomes the dominant splitting mechanism leading to the spin-split peaks in SdH oscillations. We also show that the observation of the beatings in low magnetic fields needs a system with rather higher carrier concentration so that the beatings persist up to sufficiently large fields where the oscillations are not smoothed out by Landau level broadening. The quantum Hall plateaus are evaluated and we show the Chern number with both even and odd values is replaced by the odd numbers when two subband energies are close with spin degenerate energy levels. Along with the numerical evaluation of the magnetotransport properties, a perturbative calculation is also performed which can be used in the case of low densities and not very large filling factors.
\end{abstract}


\maketitle

\section{Introduction}

The two-dimensional electron gas (2DEG) formed at the interface of two band 
insulators $\rm{LaAlO_3}$ and $\rm{SrTiO_3}$ has been the subject of various 
research topics such as metal-insulator phase transition~\cite{thiel}, tunable 
spin-orbit coupling~\cite{caviglia1}, magnetism~\cite{brinkman}, 
superconductivity~\cite{reyren} and their coexistence~\cite{dikin} to name 
some~\cite{Pai}. Quantum oscillations in the form of Shubnikov-de Haas (SdH) 
oscillations have also been observed in this system underlying the two 
dimensional character of the electronic 
states~\cite{caviglia2,shalom,fete,yang,xie}. Although there are several 
experimental reports on SdH oscillations in this system, we have not 
achieved yet a unique picture of the quantum transport at the  interface owing to 
variable samples studied and also different conditions under which the 
experiments have been performed. Moreover, large effective mass of the carriers 
at oxide interfaces compared to their semiconductor counterparts results in 
smaller Landau level separation and thus together with rather lower mobility of 
carriers in this case cause a challenging situation for observation of quantum 
conductance. By the same token, the realization of the quantum Hall effect (QHE) has 
been elusive in this system. The lack of relatively high mobility samples with 
sufficiently low carrier densities have hindered the observation of the QHE 
especially at lower filling factors. Therefore there have been so far a few and 
very diverse reports of observation of somehow imperfect plateaus of Hall 
conductance only in higher filling factors~\cite{xie,trier,Rout}. It is 
worthwhile mentioning that in most reported SdH experiments, the carrier 
concentration is smaller than the carrier concentration obtained from the Hall 
effect. The possible reasoning is that there may exist carriers with low 
mobility due to the diffusive scattering process and would be difficult to 
observe them in SdH experiment~\cite{Reyren}.

The host of the 2DEG residing at the interface of $\rm{LaAlO_3}$ and $\rm{SrTiO_3}$ are the $t_{2g}$ orbitals of $\rm{Ti}$ atom ~\cite{pentcheva,popovic}. Due to the confinement of the electron gas along $\hat{z}$, $d_{xy}$ orbital has a lower energy than  $d_{xz}$ and $d_{yz}$ orbitals. Furthermore, while $d_{xy}$ orbital forms an isotropic band with the same light mass in both $\hat{x}$ and $\hat{y}$, the  $d_{xz}$ and  $d_{yz}$ orbitals have a light effective mass in $\hat{x}$($\hat{y}$) and a heavy effective mass along $\hat{y}$($\hat{x}$) forming anisotropic bands. The inversion symmetry breaking along $\hat{z}$ at the interface together with the atomic spin-orbit interaction also result in a Rashba spin splitting as well as the orbital mixing of the bands~\cite{kim,khalsa}. Following a quasi-degenerate perturbation approach Zhou $et\, al$ developed an effective Hamiltonian around $\rm{\Gamma}$ point suggesting the usual $k$-linear Rashba spin-orbit coupling for $d_{xy}$ orbital but an anisotropic $k$-cubic spin-orbit interaction for $d_{xz}$ and $d_{yz}$ orbitals~\cite{zhou1}. The same theoretical results were found by several groups as well~\cite{kim,zhong,shanvas}. The possibility of the $k$-cubic Rashba spin splitting was also reported experimentally~\cite{nakamura,liang}. On the other hand, some magnetotransport experiments suggest that the SdH oscillations in this system originate from the carriers with rather heavy effective mass illustrating the dominant contribution of $d_{xz}$ and $d_{yz}$ orbitals in quantum oscillations~\cite{fete,yang}. This is attributed to the spatial extension of these orbitals deeper into the $\rm{SrTiO_3}$ (in comparison with closer-to-interface $d_{xy}$ orbital) so that they experience less scatterings due to disorders or lattice distortions and as a result these carriers have a high enough mobility to show the quantum oscillations.

Accordingly in this work, in order to find the magnetotransport properties of the system, we begin with the effective two-band Hamiltonian of anisotropic $d_{xz}$/$d_{yz}$ orbitals with $k$-cubic Rashba spin-orbit coupling~\cite{zhou1} and study the system in the presence of a magnetic field perpendicular to the interface. The Landau levels, as well as the chemical potential and density of states of the system, are found and we continue with the evaluation of the magnetotransport coefficients of the system. A rather anomalous beating pattern (due to the presence of the anisotropic $k$-cubic Rashba spin-orbit interaction in the Hamiltonian) is found in lower magnetic fields or larger filling factors. For larger magnetic fields, the Zeeman term becomes the dominant spin splitting mechanism so that the split conductivity peaks appear in longitudinal conductivity curves. The effects of the carrier density, Rashba strength, and Landau level broadening are also discussed and we show that in a system with rather lower carrier concentrations, the beatings are expected to appear in such a low magnetic field region where level broadenings due to disorder or temperature can even fully spoil the beating pattern.

The paper is organized as follows. In Sec.~II, the eigenvalues and eigenfunctions of the system under magnetic field along $\hat{z}$ are found both numerically and analytically (for weak Rashba interaction). The chemical potential and density of the system are also evaluated in this section. Next in Sec.~III, the expressions of both longitudinal and Hall conductivities derived from a linear-response-based formalism are introduced and discussed for our system. The numerical results of these two sections are presented in Sec.~IV together with the results of the de-Haas van Alphen oscillations of magnetization and finally the results are summarized in Sec.~V.

\section{Basic Formulation}

We consider the effective Hamiltonian introduced in Ref.~\onlinecite{zhou1} for  a pair of bands  which is a hybridization of $d_{xz}$ and $d_{yz}$ orbitals, subject to a magnetic field along $\hat{z}$

\begin{equation}
{\cal 
H}=\Pi^2/2m^*+\frac{\alpha_0}{2\hbar^3}\lbrace(\Pi_-^2+\Pi_+^2),
(\Pi\times\sigma)_{\hat{z}}\rbrace+\frac{1}{2}g\mu_BB\sigma_z,
\end{equation}
where $\alpha_0$ is the strength of the cubic spin-orbit interaction, 
$\Pi=\mathbf{p}+e\mathbf{A}$ is the canonical momentum with $\mathbf{p}$ the 
momentum operator and $\mathbf{A}$ the vector potential, $\Pi_{\pm}=\Pi_x\pm 
i\Pi_y$, $g$ is the effective Zeeman factor, and $\mu_B$ is the Bohr magneton. We also define $\lbrace A,B\rbrace=(AB+BA)/2$. In 
order to find Landau levels of the system in Landau gauge $\mathbf{A}=(0,Bx,0)$, 
we introduce the creation and annihilation operators $a=\sqrt{\frac{1}{2\hbar 
m^*\omega_c}}(\Pi_y+i\Pi_x)$ and $a^\dagger=\sqrt{\frac{1}{2\hbar 
m^*\omega_c}}(\Pi_y-i\Pi_x)$ where $\omega_c=\frac{eB}{m^*}$ is the cyclotron 
frequency, $\Pi_x=p_x$ and $\Pi_y=p_y+m^*\omega_cx$ and $m^*$ is the band 
effective mass. Rewriting the Hamiltonian using these operators, we will have 
${\cal H}=H_0+H_1+H_2$ with

\begin{equation}
H_0=\hbar\omega_c \begin{pmatrix}
(1/2+ a^\dagger a) +\xi &0 \\ 0 &(1/2+ a^\dagger a) -\xi
\end{pmatrix},
\end{equation}

\begin{equation}
H_1=\alpha \begin{pmatrix}
0 & a^\dagger a a^\dagger \\ a a^\dagger a &0
\end{pmatrix},
\end{equation}

\begin{equation}
H_2=\alpha \begin{pmatrix}
0 &a^3 \\ { a^\dagger} ^3 &0
\end{pmatrix},
\end{equation}
where $\xi=g\mu_Bm^*/2e\hbar=gm^*/4m_e$,
$\alpha=\frac{\alpha_0}{2\hbar^3}(2\hbar\omega_cm^*)^{3/2}$ and $m_e$ is the 
electron rest mass. The general way to solve eigenvalue equation  $({\cal 
H}-E I)\Psi=0$ is to expand the eigenvectors of the system in 
harmonic oscillator functions as

\begin{equation}
\begin{split}
\Psi &=\frac{e^{ik_yy}}{\sqrt{L_y}}\binom{\sum_n C_n^\uparrow \phi_n(x-x_c)}{\sum_n C_n^\downarrow  \phi_n(x-x_c)}\\
&=\frac{e^{ik_yy}}{\sqrt{L_y}}\sum_n[ C_n^\uparrow 
\phi_n(x-x_c)\ket{\uparrow}+C_n^\downarrow \phi_n(x-x_c)\ket{\downarrow}],
\end{split}
\end{equation}
Here $n$ is the Landau level index, $L_y$ is the length of the system along the $y$ direction, $\phi_n(x-x_c)=e^{-(x-x_c)^2/2l_c^2} H_n[(x-x_c)/l_c]/\sqrt{\sqrt{\pi}l_c 2^n n!}$ is the harmonic oscillator state with $ H_n[(x-x_c)/l_c]$ the Hermite polynomial of order $n$, $l_c=\sqrt{\hbar/m^*\omega_c}$ is the magnetic length, $x_c=l_c^2k_y$ is the center of oscillation and $\ket{\uparrow}$ ($\ket{\downarrow}$) is the spinor related to spin up (down).

To get more insight to the problem, it would be useful to look at cases when we have only $H_1$ or $H_2$ term.
It is interesting to note that $H_1$ is very similar to linear Rashba spin-orbit interaction~\cite{wang,zarea2005}, because it also couples two Landau levels of the order $n$ and $n-1$ with opposite spins. Solving eigenvalue problem for ${\cal H}=H_0+H_1$, the Landau levels would be

\begin{equation}
E_n^s=\hbar\omega_cn+s\hbar\omega_c(1/2+\xi)\sqrt{1+\frac{4\alpha^2n^3}{
\hbar^2\omega_c^2(1+2\xi)^2}},
\end{equation}
with $s=+1$ for spin up and $s=-1$  for spin down. The eigenfunctions are given by
\begin{eqnarray}
\psi_n^+&&=\frac{e^{ik_yy}}{\sqrt{L_y}}[\cos\theta_n\thickspace\phi_{n}\ket{
\uparrow}+\sin\theta_n\thickspace\phi_{n-1}\ket{\downarrow}], \\
\psi_n^-&&=\frac{e^{ik_yy}}{\sqrt{L_y}}[-\sin\theta_n\thickspace\phi_{n}\ket{
\uparrow}+\cos\theta_n\thickspace\phi_{n-1}\ket{\downarrow}],
\end{eqnarray}
for $n=1,2,3,\dots$. We have defined $\theta_n $ such that 
$\tan(\theta_n)=2\alpha n^{3/2}/(\delta+\delta\sqrt{1+4\alpha^2n^3/\delta^2})$,
where $\delta=\hbar\omega_c(1+2\xi)$. It can be seen that $H_1$ couples state 
$\phi_{n}\ket{\uparrow}$ with $\phi_{n-1}\ket{\downarrow}$. The only state that 
remains unchanged is the lowest spin up Landau level 
$\psi_0^+=\frac{e^{ik_yy}}{\sqrt{L_y}}\phi_{0}\ket{\uparrow}$ with eigenvalue $
E_0^+=\hbar\omega_c(1/2+\xi)$.

On the other hand, ${\cal H}=H_0+H_2$ is exactly the Hamiltonian of a system 
with isotropic cubic Rashba interaction which had been suggested for hole gas 
systems~\cite{zarea2006,mawrie}. 
In this case, the eigenvalues are
\begin{equation}
E_n^s=\hbar\omega_c(n-1)+s\hbar\omega_c(3/2-\xi)\sqrt{1+\frac{
\alpha^2n(n-1)(n-2)}{\hbar^2\omega_c^2(3/2-\xi)^2}},
\end{equation}
with eigenfunctions
\begin{eqnarray}
\psi_n^+&&=\frac{e^{ik_yy}}{\sqrt{L_y}}[\sin\theta_n\thickspace\phi_{n-3}\ket{
\uparrow}+\cos\theta_n\thickspace\phi_{n}\ket{\downarrow}], \\
\psi_n^-&&=\frac{e^{ik_yy}}{\sqrt{L_y}}[\cos\theta_n\thickspace\phi_{n-3}\ket{
\uparrow}-\sin\theta_n\thickspace\phi_{n}\ket{\downarrow}],
\end{eqnarray}
for $n=3,4,5,\dots$, where 
$\tan(\theta_n)=\alpha\sqrt{n(n-1)(n-2)}/(\delta+\delta\sqrt{
1+\alpha^2n(n-1)(n-2)/\delta^2})$ and $\delta=\hbar\omega_c(3/2-\xi)$. It is 
clear that $H_2$ mixes every $\phi_{n}\ket{\uparrow}$ with 
$\phi_{n+3}\ket{\downarrow}$. Again, we have three states $n=0,1,2$, which do 
not couple with other states, i.e. eigenvalues 
$E_n^-=\hbar\omega_c(n+1/2)-\hbar\omega_c\xi$ and eigenfunctions 
$\psi_n^-=\frac{e^{ik_yy}}{\sqrt{L_y}}\phi_{n}\ket{\downarrow}$. Therefore, for 
$n=0,1,2$ we do not have any $+$ branch.

When both $H_1$ and $H_2$ are present, the Hamiltonian should be diagonalized
numerically in a truncated Hilbert space with the sufficiently large number of basis functions
. We can see that in this case, the Landau levels of the system fall
into the following four groups with different couplings
\begin{eqnarray*}
&\phi_{0}\ket{\downarrow}\xrightarrow{H_1}\phi_{1}\ket{\uparrow}\xrightarrow{H_2
}\phi_{4}\ket{\downarrow}\xrightarrow{H_1}\phi_{5}\ket{\uparrow}\xrightarrow{H_2
}+\dotsb\thickspace,  (I)\\
&\phi_{1}\ket{\downarrow}\xrightarrow{H_1}\phi_{2}\ket{\uparrow}\xrightarrow{H_2
}\phi_{5}\ket{\downarrow}\xrightarrow{H_1}\phi_{6}\ket{\uparrow}\xrightarrow{H_2
}+\dotsb\thickspace,  (II)\\
&\phi_{2}\ket{\downarrow}\xrightarrow{H_1}\phi_{3}\ket{\uparrow}\xrightarrow{H_2
}\phi_{6}\ket{\downarrow}\xrightarrow{H_1}\phi_{7}\ket{\uparrow}\xrightarrow{H_2
}+\dotsb\thickspace,  (III)\\
&\phi_{0}\ket{\uparrow}\xrightarrow{H_2}\phi_{3}\ket{\downarrow}\xrightarrow{H_1
}\phi_{4}\ket{\uparrow}\xrightarrow{H_2}\phi_{7}\ket{\downarrow}\xrightarrow{H_1
}+\dotsb\thickspace,  (IV)\\
\end{eqnarray*}

On the other hand, because of the large value of the Zeeman term with respect to 
the Rashba interaction, we can find an approximate solution to this problem treating 
both $H_1$ and $H_2$ as perturbations to $H_0$. Therefore, to the second order 
of 
perturbations the Landau levels are
\begin{eqnarray}
E_n^+&&={E_n^0}^{+}+\frac{\alpha^2n^3}{(1+2\xi)\hbar\omega_c}+\frac{
\alpha^2(n+3)(n+2)(n+1)}{(-3+2\xi)\hbar\omega_c},\nonumber \\ 
\label{eq:eig1}\\
E_n^-&&={E_n^0}^{-}-\frac{\alpha^2(n+1)^3}{(1+2\xi)\hbar\omega_c}-\frac{
\alpha^2n(n-1)(n-2)}{(-3+2\xi)\hbar\omega_c},
\label{eq:eig2}
\end{eqnarray}
where ${E_n^0}^{\pm}=(n+1/2\pm\xi)\hbar\omega_c$. 

\begin{figure}[ht]
\centering
      \includegraphics[width=1.\linewidth] {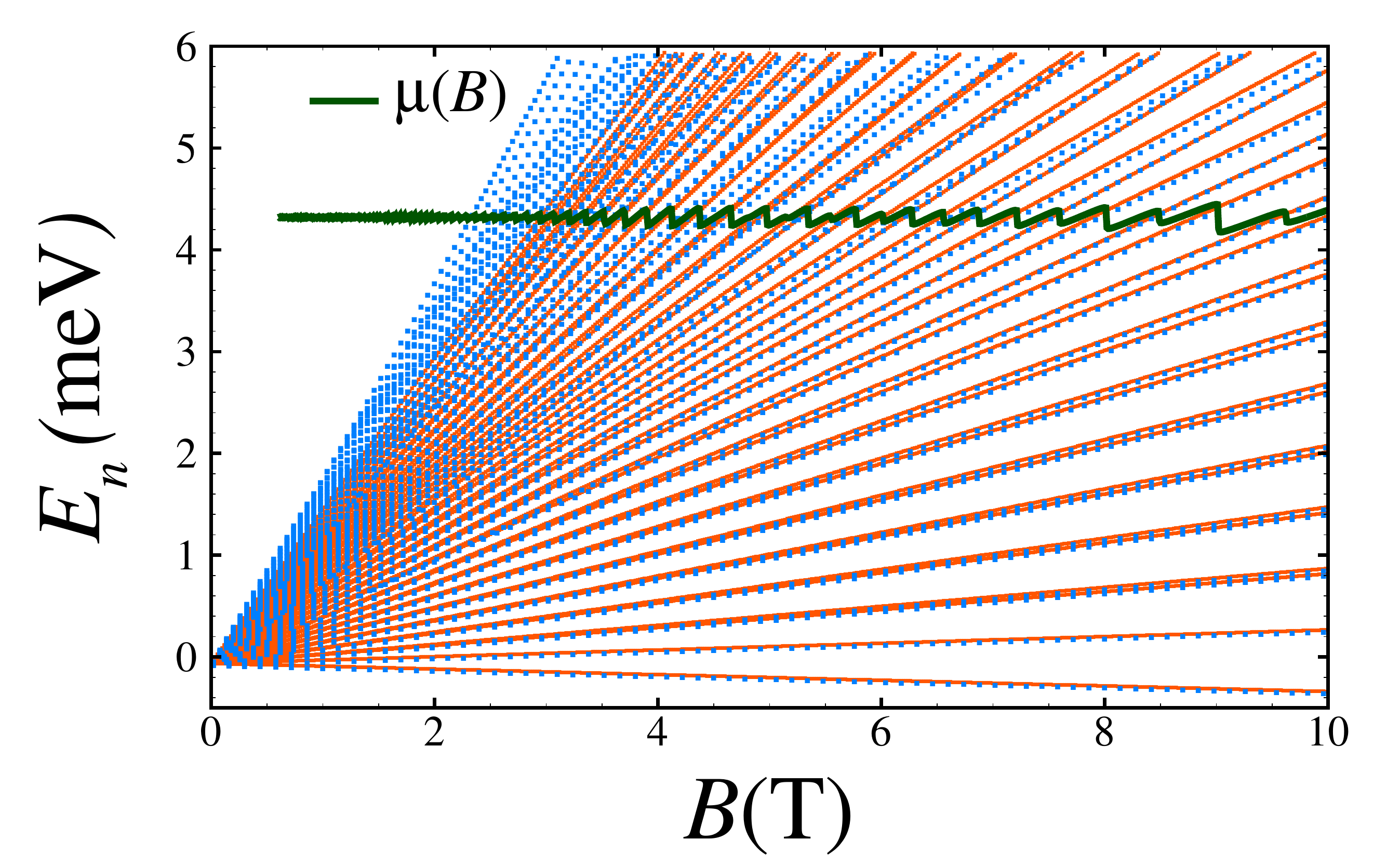}
\caption{\label{fig1} (Color online)  Landau levels of the system evaluated 
numerically (solid lines) in comparison with those found using  
Eqs.~\eqref{eq:eig1} and \eqref{eq:eig2} (dotted lines) versus magnetic field 
$B$.  The perturbative Landau levels provided by Eqs. \eqref{eq:eig1} and 
\eqref{eq:eig2} are in good agreement with the exact numeric ones especially 
for the lower Landau levels and higher magnetic fields. The oscillating chemical 
potential is also shown for $n_c=\rm{3.5\times 10^{16}~m^{-2}}$ showing that the 
chemical potential depends weakly on $B$ . Here, we use $g=2$, 
$\alpha_0=5.5\,\rm{eV\AA^3}$, $m^*=1.9\, m_e$, $T=100\, \rm{mK}$.    }
\end{figure}

In Fig. \ref{fig1} we depict 
the Landau levels of the system obtained numerically (solid lines) in 
comparison 
to thouse found from  Eqs.~\eqref{eq:eig1} and \eqref{eq:eig2} (dashed lines). We 
can see that the perturbative Landau levels are in good agreement with the 
exact ones especially for lower Landau levels and higher magnetic fields. The 
eigenfunctions, in the first order of perturbation, have also been found as
\begin{equation}\label{eq:psi_+}
\psi_n^+=\frac{e^{ik_yy}}{\sqrt{L_y}}N_{n,+}[\phi_{n}\ket{\uparrow}+C_{n,+}\phi_{
n-1}\ket{\downarrow}+D_{n,+}\phi_{n+3}\ket{\downarrow}],
\end{equation}
\begin{equation}\label{eq:psi_-}
\psi_n^-=\frac{e^{ik_yy}}{\sqrt{L_y}}N_{n,-}[\phi_{n}\ket{\downarrow}+C_{n,-}
\phi_{n+1}\ket{\uparrow}+D_{n,-}\phi_{n-3}\ket{\uparrow}],
\end{equation}
where $N_{n,\pm}=1/\sqrt{1+C_{n,\pm}^2+D_{n,\pm}^2}$,
\begin{eqnarray}
C_{n,+}&&=\alpha n^{3/2}/(1+2\xi)\hbar\omega_c, \nonumber \\
D_{n,+}&&=\big(\frac{\alpha\sqrt{(n+1)(n+2)(n+3)}}{(-3+2\xi)\hbar\omega_c}\big){
\Theta } (n-1) , \nonumber \\
C_{n,-}&&=-\alpha (n+1)^{3/2}/(1+2\xi)\hbar\omega_c, \nonumber \\
D_{n,-}&&=\big(\frac{-\alpha\sqrt{n(n-1)(n-2)}}{(-3+2\xi)\hbar\omega_c}\big)
{\Theta } (n-3) , \nonumber 
\end{eqnarray}
and 
$\Theta(x)$ is the Heaviside function.

Knowing the electron concentration of the system, we can find the field 
dependent chemical potential of the system through
\begin{equation}\label{eq:nc}
n_c=\int_{-\infty}^{\mu(B)}D(\varepsilon)f(\varepsilon)d\varepsilon,
\end{equation}
where $f(\varepsilon)=(1+\exp[\beta(\varepsilon-\mu(B))])^{-1}$ is the Fermi-Dirac distribution and $\beta=1/k_BT$. The density of states is 
defined by
\begin{equation}\label{eq:D_E}
D(\varepsilon)=1/A_0\sum_{n,s,k_y}\delta(\varepsilon-E_n^s),
\end{equation}
with $A_0$ is the area of the system. For the sum over $k_y$, we get 
$\sum_{k_y}\rightarrow g_s L_y/2\pi \int_{-L_x/2l_c^2}^{L_x/2l_c^2} dk_y=g_s 
A_0/2\pi l_c^2$. Therefore, Eq.~\eqref{eq:nc} is given by $n_c=1/2\pi l_c^2 
\sum_{n,s} f(E_n^s)$, ($g_s$=1 since the spin degeneracy is lifted in our 
problem).

Assuming a Landau level broadening of the width $\Gamma$~\cite{fete,yang}, we can 
replace the Delta function of Eq.~\eqref{eq:D_E} with a Gaussian and find the 
density of states as
\begin{equation}
D(\varepsilon)=\frac{g_s}{D_0\sqrt{2\pi}\Gamma}\sum_{n,s}\exp[\frac{
-(\varepsilon-E_n^s)^2}{2\Gamma^2}],
\end{equation}
where $D_0=2\pi l_c^2$. We will discuss the chemical potential and density of states of the system in more details in  Sec.~IV.

\section{Magnetotransport Coefficients}

In order to find the magnetotransport properties of the system, we follow the
approach based on the linear response formalism (as well as the quantum
Boltzmann equation) introduced and developed by Van Vilet $et\, al$ in a series
of papers~\cite{vilet1978,vilet1979,charbonneau,vasilopoulos1984} and then
widely applied to 2DEGs in the presence of a perpendicular magnetic field. In this approach, the
full Hamiltonian of the system is considered to be composed of a diagonal part,
which is the largest part of the Hamiltonian, a nondiagonal interaction part
(such as the interaction between electron-impurity or electron-phonon) and
finally an external field part, which is the electric field in this case. As a
result, the conductivity tensor of the system will have a diagonal as well as a
nondiagonal part $\sigma_{\mu\nu}=\sigma_{\mu\nu}^d+\sigma_{\mu\nu}^{nd}$ with
$\sigma_{\mu\nu}^d=\sigma_{\mu\nu}^{dif}+\sigma_{\mu\nu}^{col}$. The first term
is the usual diffusive current while the second term is the collisional
conductivity, which is the many-body contributions of collisions. The diffusive
conductivity contains the diagonal elements of the velocity operator, which is
zero for both the longitudinal and Hall conductivities. On the other hand,
$\sigma_{xx}^{nd}$ and $\sigma_{yx}^{col}$ also vanish~\cite{charbonneau} so that the only terms
contributing to the longitudinal and Hall conductivities are
$\sigma_{xx}=\sigma_{xx}^{col}$ and $\sigma_{yx}=\sigma_{yx}^{nd}$.

Therefore, the longitudinal conductivity is given by
\begin{equation}\label{eq:sigma_xx0}
\begin{split}
\sigma_{xx}^{col}=&\frac{\beta e^2}{A_0}\sum_{\zeta\zeta'}\int d\varepsilon \int d\varepsilon' \delta(\varepsilon-E_{\zeta})\delta(\varepsilon'-E_{\zeta'}) \\
&\times 
f(\varepsilon)[1-f(\varepsilon')]W_{\zeta\zeta'}(x_{\zeta}-x_{\zeta'})^2,
\end{split}
\end{equation}
where $\ket{\zeta}=\ket{n,s,k_y}$ and $x_{\zeta}=\bra{\zeta}x\ket{\zeta}$. Considering elastic scattering between carriers and impurities with screened potential of the form $U(\mathbf{r})=e^2 e^{-k_sr}/4\pi\epsilon_0\epsilon r$, we can define the transition rate between state $\ket{\zeta}$ and $\ket{\zeta'}$ as
\begin{equation}
W_{\zeta\zeta'}=\frac{2\pi n_i}{\hbar A_0}\sum_q 
|U(\mathbf{q})|^2|F_{\zeta\zeta'}(\mathbf{q})|^2\delta(\varepsilon-\varepsilon')
,
\end{equation}
where $n_i$ is the impurity density, $F_{\zeta\zeta'}=\bra{\zeta}e^{i\mathbf{q}.\mathbf{r}}\ket{\zeta'}$ is the form factor and $U(\mathbf{q})=e^2/[2\epsilon_0\epsilon(q^2+k_s^2)^{1/2}]$ is the Fourier transform of the impurity potential. Here $\epsilon_0$ is the vacuum permittivity, $\epsilon$ is the dielectric constant and $k_s$ is the screening wave vector. Since the many-body correlations are suppressed in the presence of the magnetic field, we therefore, consider the Hartree-Fock screening of the system. In order to calculate $\sigma_{xx}^{col}$, we should note that the evaluation of $F_{\zeta\zeta'}$ leads to $\delta_{k'_y,k_y-q_y}$ and therefore we have $(x_{\zeta}-x_{\zeta'})^2=(k_y-k'_y)^2l_c^4=q^2\sin^2\phi l_c^4$.  We also have $\sum_{k_y}=A_0/2\pi l_c^2$ and $\sum_{q}\rightarrow A_0/(2\pi)^2 \int qdqd\phi$. We can once again replace the Delta functions in \eqref{eq:sigma_xx0} with a Gaussian to account for level broadening. Performing the integration over $\phi$ and setting $u=q^2l_c^2/2$, we finally have
\begin{equation}\label{eq:sigma_xx}
\begin{split}
\sigma_{xx}^{col}=&\frac{ e^2}{h}\frac{\beta}{2\pi\Gamma^2l_c^2}n_i U_0^2 \sum_{\zeta}\int duu |F_{\zeta}(u)|^2\\
&\times\int d\varepsilon\exp[\frac{-(\varepsilon-E_\zeta)^2}{\Gamma^2}] 
f(\varepsilon)[1-f(\varepsilon)],
\end{split}
\end{equation}
with $F_{\zeta}(u)\equiv F_{\zeta\zeta}(u)$ and $U_0=e^2/2\epsilon_0\epsilon 
k_s$. The above expression is found using the fact that 
$|F_{\zeta}(u)|^2\sim e^{-u}$(see Appendix.~\ref{ap-one}), the most important 
contribution to the integral comes from small $u$ and therefore in the 
denominator of $U(\mathbf{q})$ we can neglect $q^2$ with respect to $k_s$ . To 
evaluate the above expression, we should calculate the integrals of the form 
$I_{\zeta}=\int u |F_{\zeta}(u)|^2 du$. In an appropriate condition discussed 
before, this integral can be evaluated analytically using the approximate 
eigenvalues and eigenfunctions found in previous section (see
Appendix.~\ref{ap-one}). We can estimate the level broadening from 
$\Gamma_{\zeta}=\hbar\sum_{\zeta'}W_{\zeta\zeta'}$ without considering the 
spin-orbit interaction~\cite{vasilopoulos1986} so that we have $\Gamma^2\approx 
n_iU_0^2/\sqrt{2\pi}l_c^2$. We restore the $U_0$ and $k_s$ in the parameter 
$\Gamma$.

It should be noted that in general case of the electronic mobility of LAO/STO
interface, an anisotropic three band Boltzmann equation is needed to find the transport properties of the system~\cite{Faridi}. To do so, a three band Hamiltonian with the anisotropic characteristic of the dielectric function might be considered. 

Since the time-reversal symmetry is broken, the hall conductivity is nonzero. The Hall conductivity can thus be written as
\begin{equation}\label{eq:sigma_yx}
\sigma_{yx}^{nd}=\frac{i\hbar 
e^2}{A_0}\sum_{\zeta\neq\zeta'}\frac{f(E_\zeta)-f(E_{\zeta'})}{(E_\zeta-E_{
\zeta'})(E_\zeta-E_{\zeta'}+i\Gamma_\zeta)}\upsilon_{\zeta\zeta'}^y 
\upsilon_{\zeta'\zeta}^x,
\end{equation}
where $\upsilon_{\zeta\zeta'}^y=\bra{\zeta}\upsilon_y\ket{\zeta'}$ and  $\upsilon_{\zeta'\zeta}^x=\bra{\zeta'}\upsilon_x\ket{\zeta}$  are the non-diagonal matrix element of the velocity operator and we assume that $\Gamma_\zeta=\Gamma$. In order to obtain an expression for the Hall conductivity, we need to have the components of the velocity operator $\upsilon_x=\partial H/\partial p_x$ and $\upsilon_y=\partial H/\partial p_y$ which read as
\begin{equation}
\upsilon_x\!=\! \begin{pmatrix}
i\gamma (a^\dagger -a) &i\lambda(3a^2\!+\!{a^\dagger}^2\!-\!2a^\dagger a\!-\!1) \\ -i\lambda(3{a^\dagger}^2\!+\!a^2\!-\!2a^\dagger a\!-\!1)&i\gamma (a^\dagger -a)
\end{pmatrix},
\end{equation}

\begin{equation}
\upsilon_y\!=\! \begin{pmatrix}
\gamma (a^\dagger+a) &\lambda(3a^2\!+\!{a^\dagger}^2\!+\!2a^\dagger a\!+\!1) \\ -\lambda(3{a^\dagger}^2\!+\!a^2\!+\!2a^\dagger a\!+\!1)&\gamma (a^\dagger +a)
\end{pmatrix},
\end{equation}
with $\gamma=\sqrt{2\hbar m^*\omega_c}/2$ and 
$\lambda=\alpha_0m^*\omega_c/\hbar^2$. 
The matrix elements of the velocity operator are given in
Appendix.~\ref{ap-two} for the perturbative regime. We see that the summation 
over $k_y$ in Eq.~\eqref{eq:sigma_yx} gives $A_0/2\pi l_c^2$ and also the 
Kronecker delta appearing in the matrix elements of the velocity operator allows 
only special values for $n'$ , so considering also $\delta_{k_y,k'_y}$ we have 
$\Sigma_{\zeta\neq\zeta'}\rightarrow 
\Sigma_{n,++}+\Sigma_{n,--}+\Sigma_{n,+-}+\Sigma_{n,-+}$, which show both 
intrabranch scatterings the same as interbranch ones.

The resistivity tensor can also be written in terms of the conductivity tensor such that for magnetoresistivity we have $\rho_{xx}=\sigma_{yy}/(\sigma_{xx}\sigma_{yy}-\sigma_{xy}\sigma_{yx})$ and the Hall resistivity is $\rho_{yx}=-\sigma_{yx}/(\sigma_{xx}\sigma_{yy}-\sigma_{xy}\sigma_{yx})$.

\begin{figure}[h]
\centering
\subfloat{%
  \includegraphics[width=1.\linewidth]{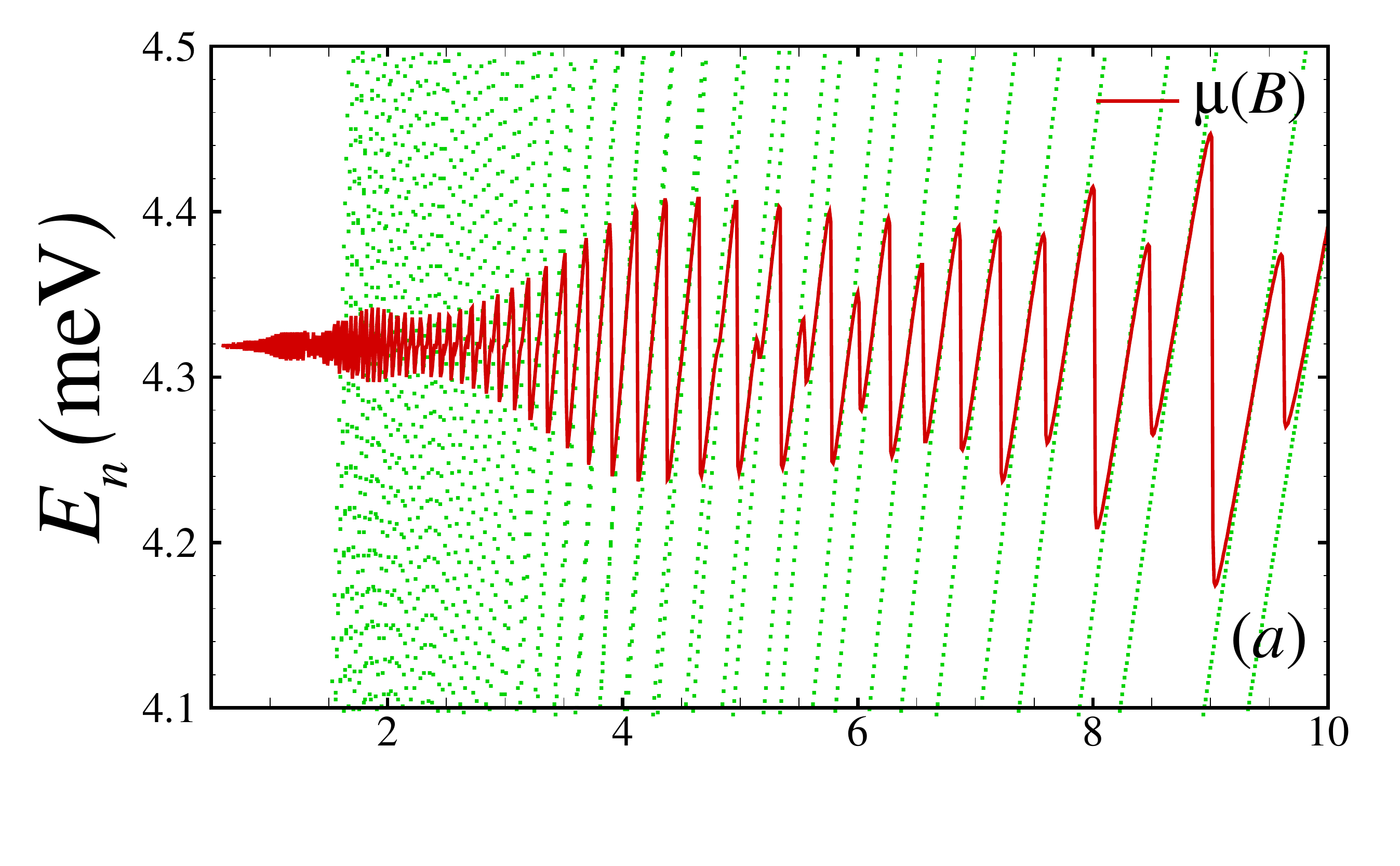}%
  }\par 
\negthickspace\negthickspace\negthickspace\negthickspace\negthickspace\negthickspace\negthickspace\negthickspace\negthickspace\negthickspace\negthickspace\negthickspace
\subfloat{%
  \includegraphics[width=1.\linewidth]{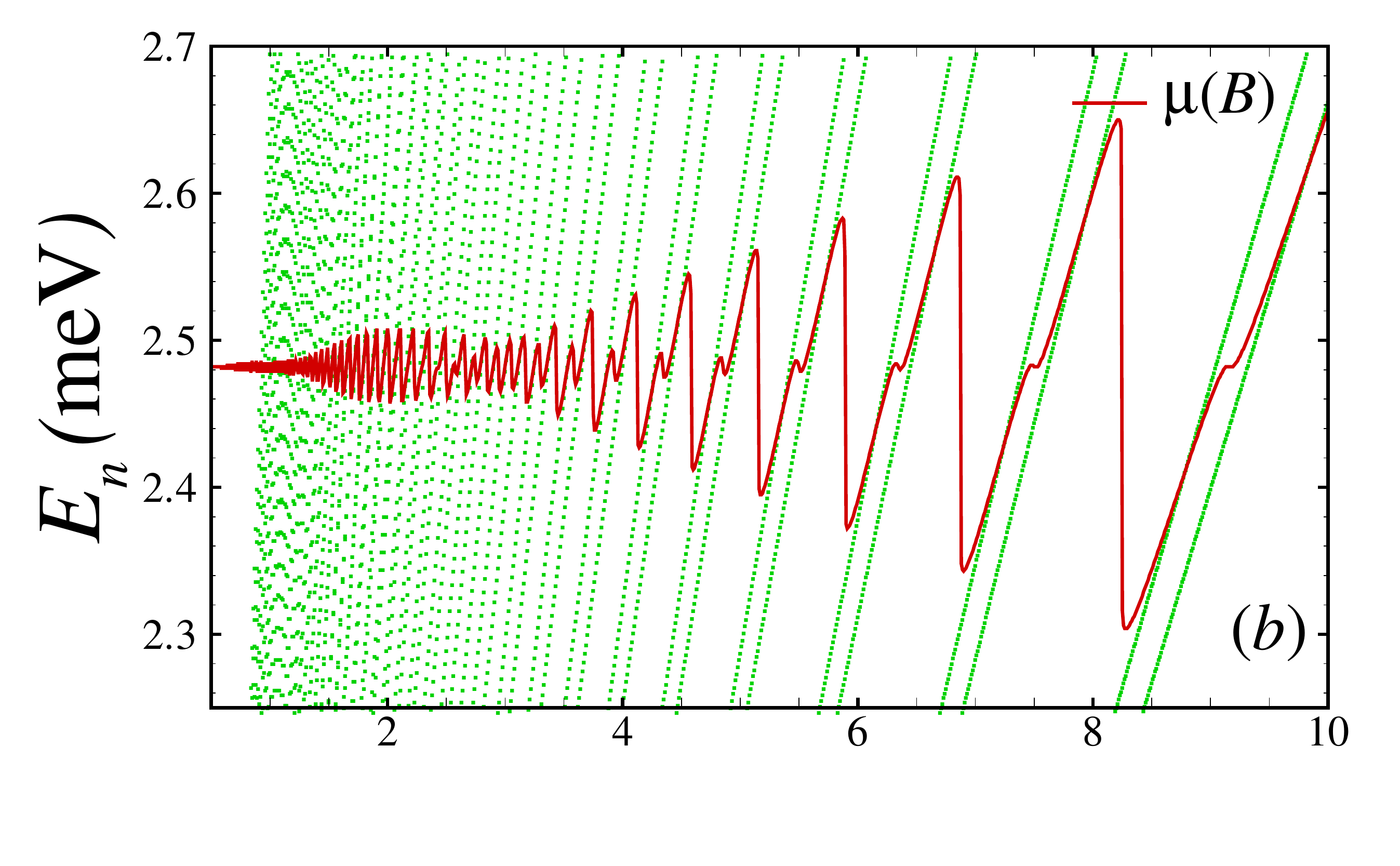}%
  }\par 
\negthickspace\negthickspace\negthickspace\negthickspace\negthickspace\negthickspace\negthickspace\negthickspace\negthickspace\negthickspace\negthickspace\negthickspace
\subfloat{%
  \includegraphics[width=1.\linewidth]{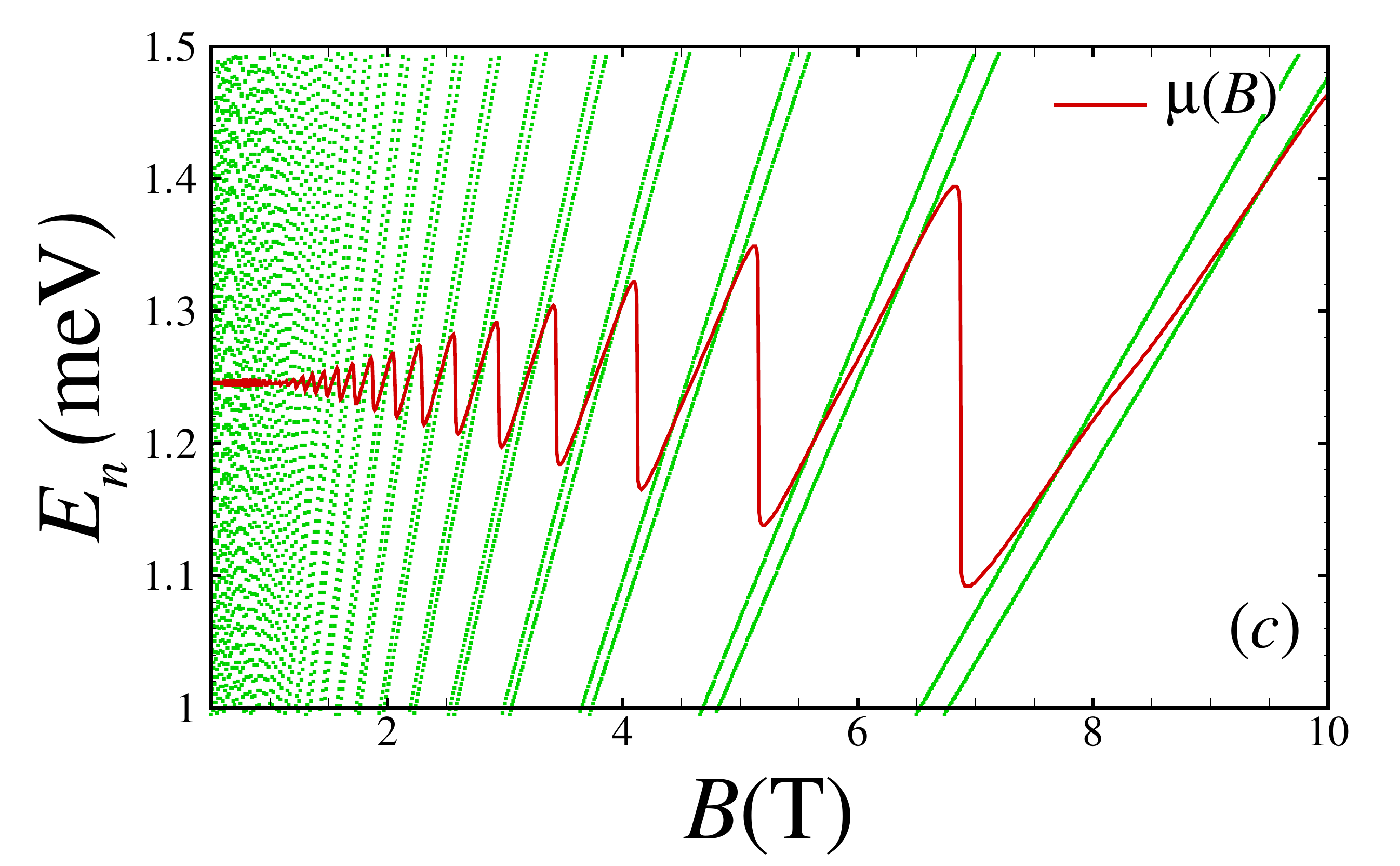}%
  }
\caption{\label{fig2} (Color online)  Landau levels (dotted lines) and the chemical potential (solid line) of the system at  $T=100\, \rm{mK}$ as functions of magnetic field $B$ for $(a)\,n_c=\rm{3.5\times 10^{16}~m^{-2}}$, $(b)\,n_c=\rm{2\times 10^{16}~m^{-2}}$ and $(c)\,n_c=\rm{1\times 10^{16}~m^{-2}}$.  The chemical potential shows a beating pattern in lower magnetic fields specially for higher densities.
    }
\end{figure}

\section{Results and Discussions}

In this section, we present the numerical results obtained for the 
magnetotransport properties of a two-band 2DEG with an anisotropic cubic Rashba 
spin-orbit interaction. An anomalous beating pattern is found in the low 
magnetic field regime for all quantities reported here involving the chemical 
potential, density of states and also longitudinal conductivity and resistivity. 
The beating pattern is the characteristic of systems with two subbands each of 
which has a different level spacing, so that the energy levels of one subband 
grows faster than that of the other one. Therefore, the energy level of one 
subband can sometimes be located in the middle of two energy levels of the other 
subband creating a node in the density profile. On the other hand, the energy levels 
of the two subbands can also meet making a maximum in quantum oscillations.

These two subbands are the Rashba spin-split bands in our case. We should note that although even in the absence of the Rashba splitting, the Zeeman term splits the bands into two branches, since the Landau level spacing of both branches is the same, no beating pattern shows up in this case. We can conclude from the perturbed energy levels that the Rashba splitting is proportional to $n^3$ ($n$ being the Landau level index), so for lower magnetic fields where more Landau levels are occupied, the significant role of Rashba spin splitting leads to the emergence of the beating pattern. As we increase the magnetic field and the higher Landau levels become empty, the constant Zeeman splitting will be the dominant spin splitting mechanism.

If not otherwise specified, the parameters we used for our calculations are; 
$g=2$~\cite{caviglia1}, $\alpha_0=5.5\,\rm{eV\AA^3}$, $m^*=1.9\, m_e$ ($m_e$
being free electron mass)~\cite{m-alpha} and $n_c=\rm{3.5\times
10^{16}~m^{-2}}$. The density concentration is chosen such that we can find an appropriate insight to the problem. The same reasoning is also valid for lower densities. To cover a large magnetic field range in our calculations, we have evaluated the magnetotransport properties of the system
numerically, since we pointed out earlier that the perturbative
expressions are applicable for lower density and higher magnetic field regimes.

\begin{figure}[h]
\centering
\subfloat{%
  \includegraphics[width=1.\linewidth]{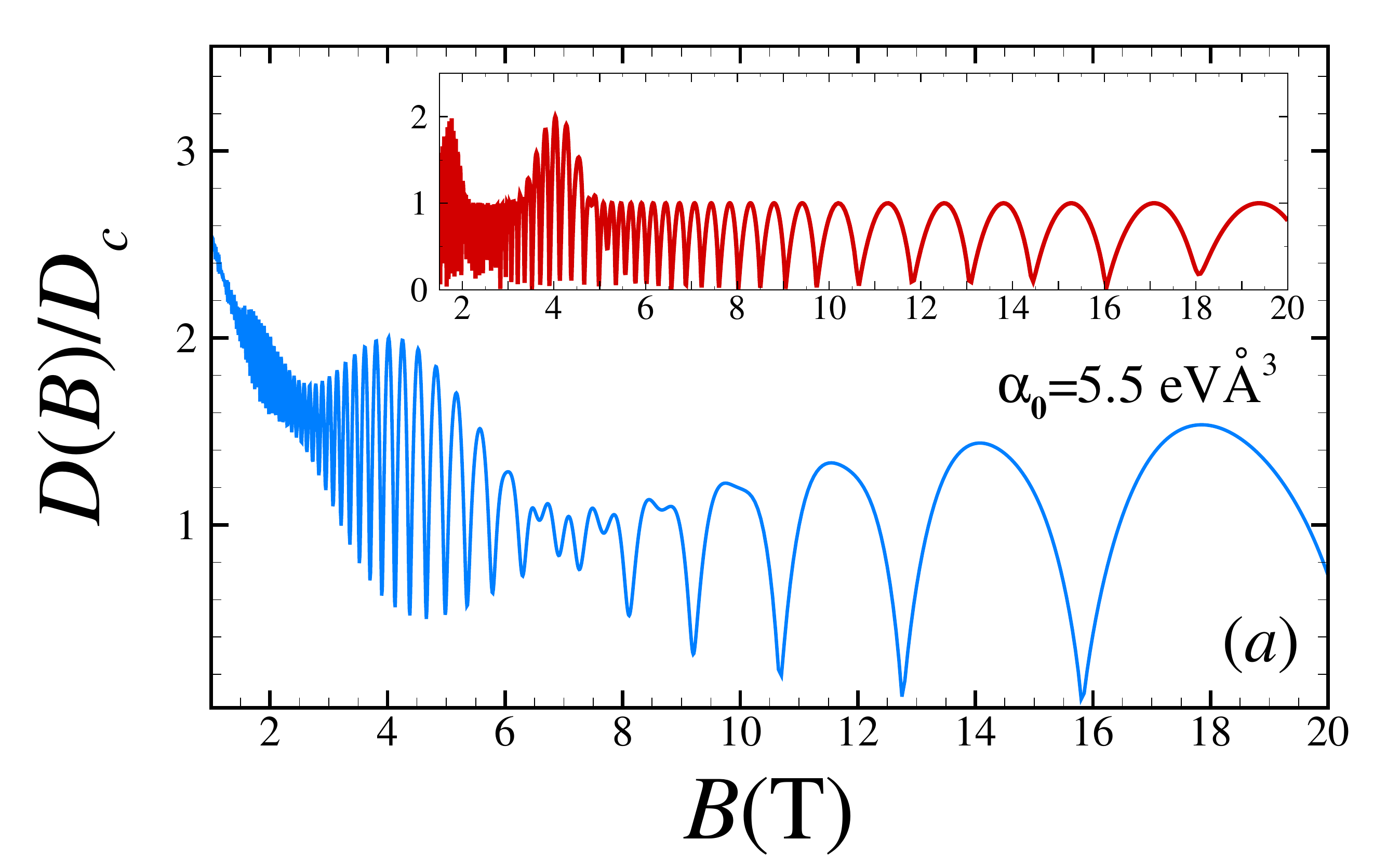}%
  }\par 
\negthickspace\negthickspace\negthickspace\negthickspace\negthickspace\negthickspace\negthickspace\negthickspace\negthickspace\negthickspace\negthickspace\negthickspace
\subfloat{%
  \includegraphics[width=1.\linewidth]{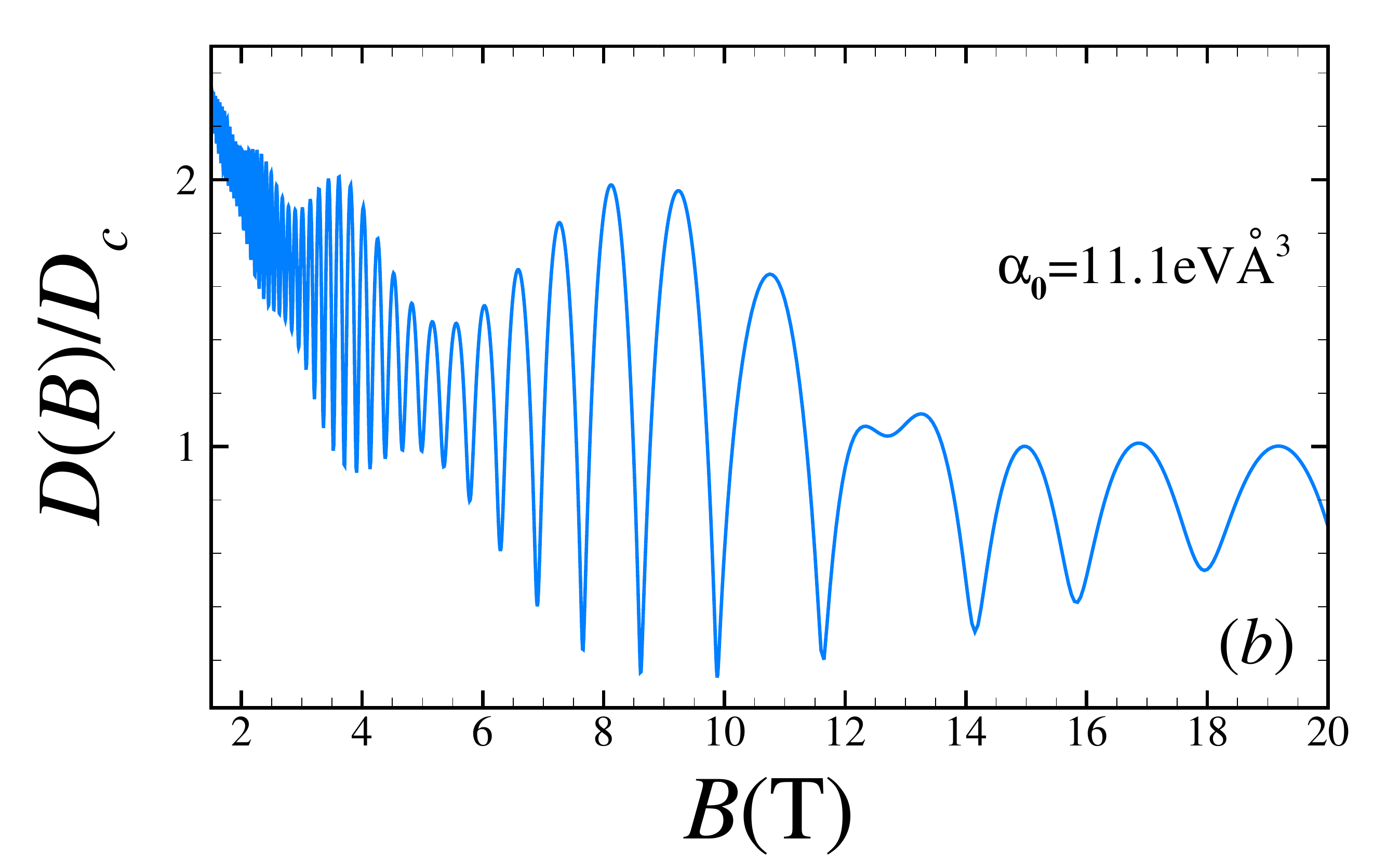}%
  }
\caption{\label{fig3} (Color online)  Dimensionless density of states of the system at  $T=100\, \rm{mK}$ as a function of magnetic field $B$ for $(a)\, \alpha_0=5.5\,\rm{eV\AA^3}$ and $(b)\,\alpha_0=11.1\,\rm{eV\AA^3}$. Both figures are plotted for $\Gamma=0.03\sqrt{B}\, \rm{meV}$. Inset: The same as in $(a)$ but with $\Gamma=0.01\sqrt{B}\, \rm{meV}$.     }
\end{figure}

In Fig. \ref{fig2}, we show the chemical potential of the system for different density concentrations. As expected, a beating pattern can be seen in lower magnetic fields specially for higher densities. This is obvious because the Rashba spin splitting is related to $k_F$ through $\Delta_R=2\alpha k_F^3|\cos2\theta|$. Therefore, in lower densities, $\Delta_R$ is too small that the beating pattern appears only in very low magnetic fields.

In Fig. \ref{fig3}, we plot the dimensionless density of states of the system
$D(B)/D_c$  for different strengths of the Rashba interaction
($D_c=g_s/D_0\sqrt{2\pi}\Gamma$). We see that the oscillations of Fig.
\ref{fig3}(a) follow the same pattern as the chemical potential with the same density. Fig. \ref{fig3}(b) shows the same result for a larger value of $\alpha_0$. In this case, the beating pattern persists up to larger magnetic fields with more periods of beatings. As we stated before, the Rashba interaction is responsible for the foundation of the beating pattern so that a stronger Rashba interaction results in beatings, which survives up to larger magnetic fields.
The density pattern is illustrated in the inset of Fig. \ref{fig3}(a) for a
smaller Landau level broadening. We can see that in this case, the periods of the beatings are not formed completely in the low magnetic field region and the Zeeman spin-split peaks can be distinguished in larger magnetic fields. Furthermore,
the piecewise parabolic pattern of the density of states in terms of the magnetic field at large $B$, suggests a similar magnetic dependence of the ground-state energy of the system.

\begin{figure}[h]
\centering
      \includegraphics[width=1.\linewidth] {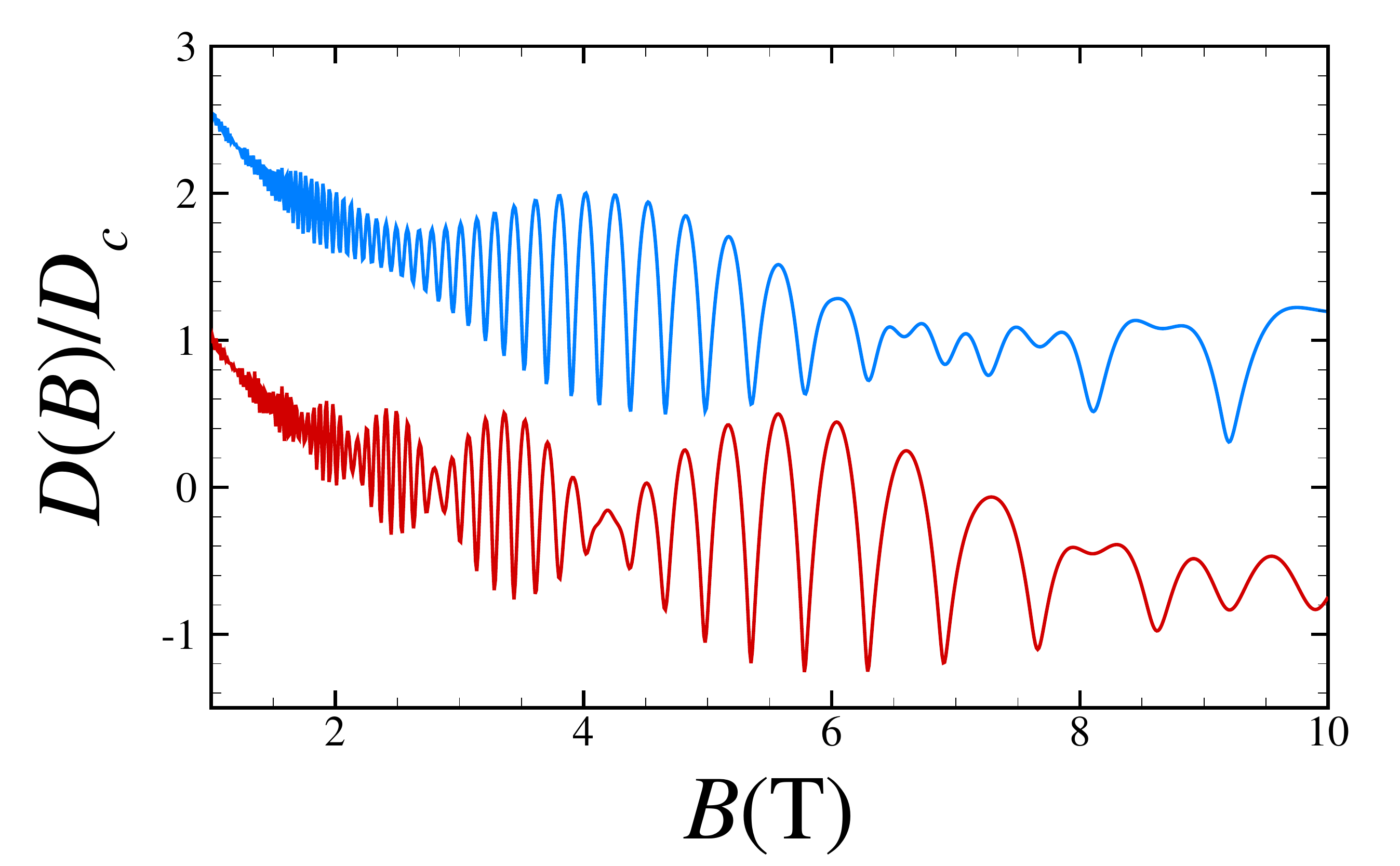}
\caption{\label{fig4} (Color online)  Dimensionless density of states of the system at  $T=100\, \rm{mK}$ as a function of the magnetic field $B$ (upper figure) in comparison with that of a system with isotropic k-cubic Rashba interaction (lower figure). Both figures are plotted for $\Gamma=0.03\sqrt{B}\, \rm{meV}$ and $ \alpha_0=5.5\,\rm{eV\AA^3}$. The lower graph is shifted vertically for clarity.    }
\end{figure}

We compare the density profile of the present system with the one with the isotropic k-cubic Rashba interaction in Fig. \ref{fig4}. The oscillations of the present system in a low magnetic field show a rather anomalous behavior with not the equal number of the peaks in each beating period in comparison with that of the isotropic Rashba one. Such anomalous beating pattern is expected in the systems with anisotropic spin splitting owing to the magnetic break down at the points on the Fermi surface with a smaller spin splitting~\cite{andrada}. In these special points, the electrons can tunnel from one of the Fermi surfaces to another one.

\begin{figure}[h]
\centering
      \includegraphics[width=1.\linewidth] {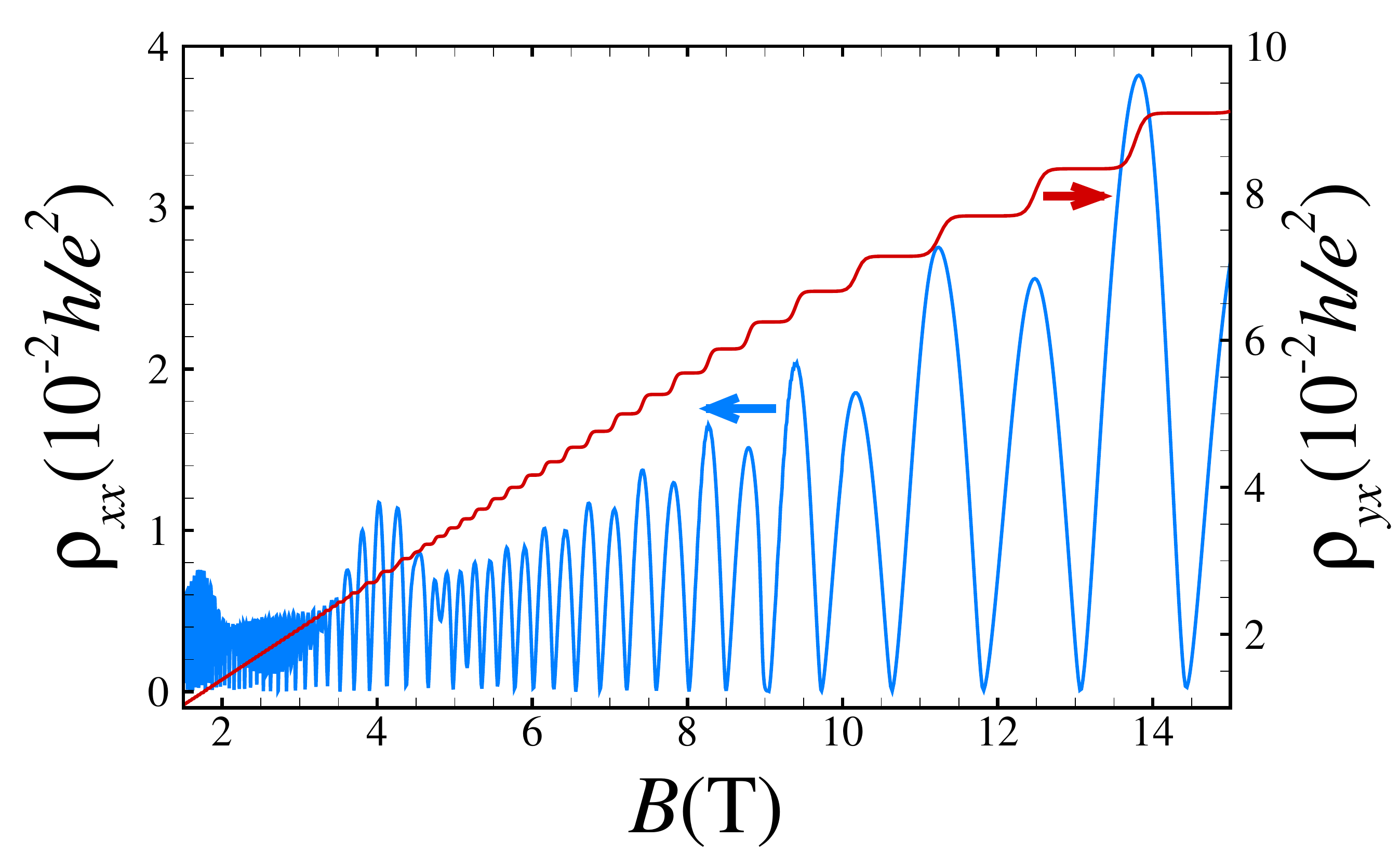}
\caption{\label{fig5} (Color online) Hall and longitudinal resistivities versus magnetic field $B$ for $T=50\, \rm{mK}$.  The steps between integer quantum Hall plateaus coincide well with the peaks of the longitudinal resistivity. For lower magnetic fields, as the beatings begin to show up in longitudinal resistivity the Hall plateaus also seem to behave in a different way.  }
\end{figure}

The longitudinal and Hall resistivities are plotted in Fig. \ref{fig5}. For evaluation of the longitudinal resistivity the Landau level broadening is assumed to be $\Gamma=0.01\sqrt{B}\, \rm{meV}$ in accordance with the experimental reports finding a much smaller value for Landau level broadening compared to their spacing~\cite{shalom,xie}. The same oscillating pattern (as in Fig. \ref{fig2}(a)) is seen in the longitudinal resistivity of the system. The somehow incomplete beatings persist up to about $B\approx4.7\,\rm{T}$. As we decrease the density concentration to for example $n_c=2\times10^{16}\rm{m^{-2}}$ the beatings disappear for magnetic fields larger than $B\approx2\,\rm{T}$. We should note that the beating pattern exists in much lower magnetic fields as well, but it is suppressed in this limit due to the Landau level broadening effect. The Hall resistivity versus magnetic field is also shown in this figure. It can be seen that the steps between integer quantum Hall plateaus (in units of $h/e^2$) coincide well with the sharp peaks of the longitudinal resistivity. For lower magnetic fields, as the beatings begin to show up in the longitudinal resistivity, the Hall plateaus also seem to behave in a different way.

\begin{figure}[h]
\centering
\subfloat{%
  \includegraphics[width=1.\linewidth]{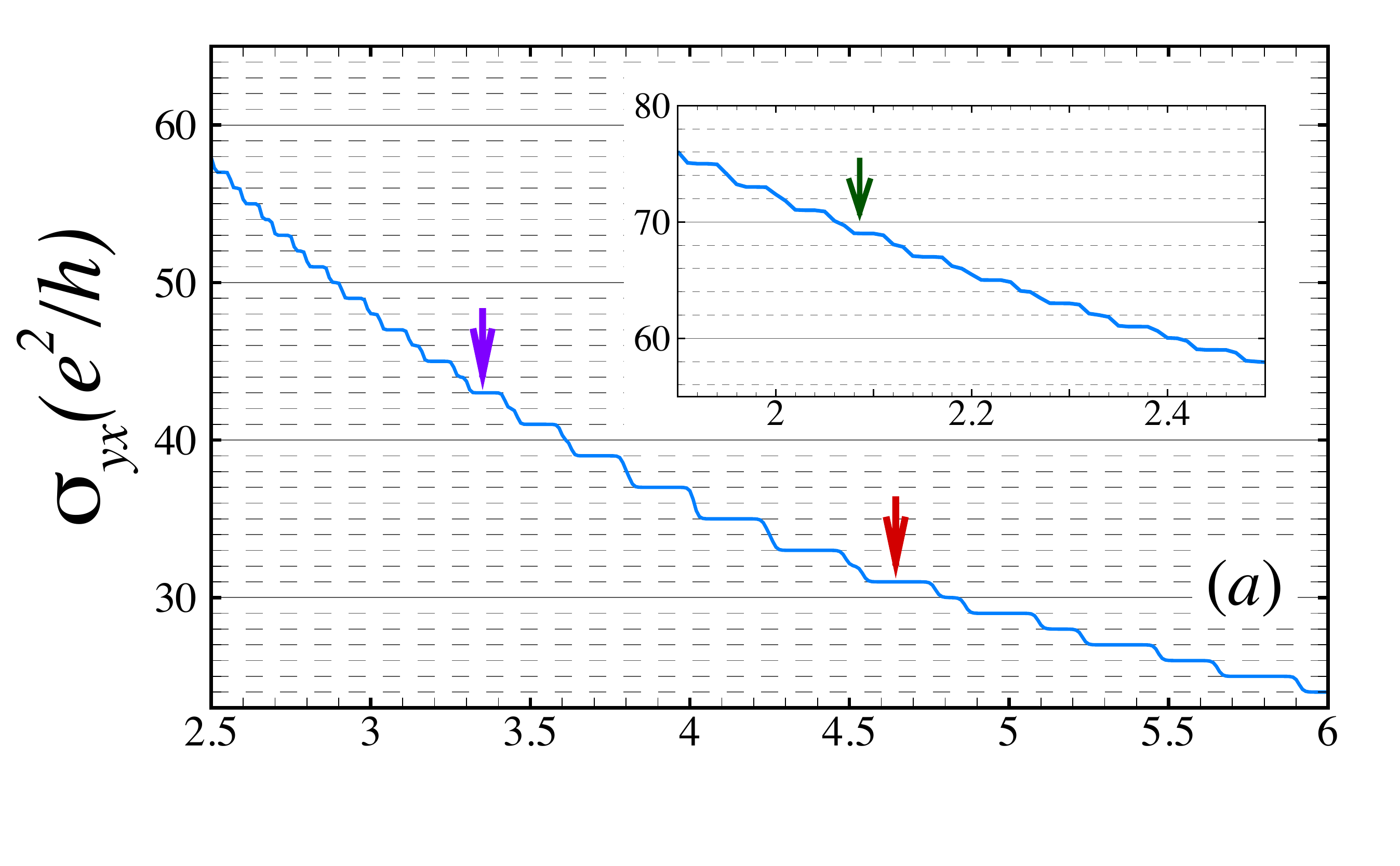}%
  }\par 
\negthickspace\negthickspace\negthickspace\negthickspace\negthickspace\negthickspace\negthickspace\negthickspace\negthickspace\negthickspace\negthickspace\negthickspace\negthickspace
       
\subfloat{%
  \includegraphics[width=1.\linewidth]{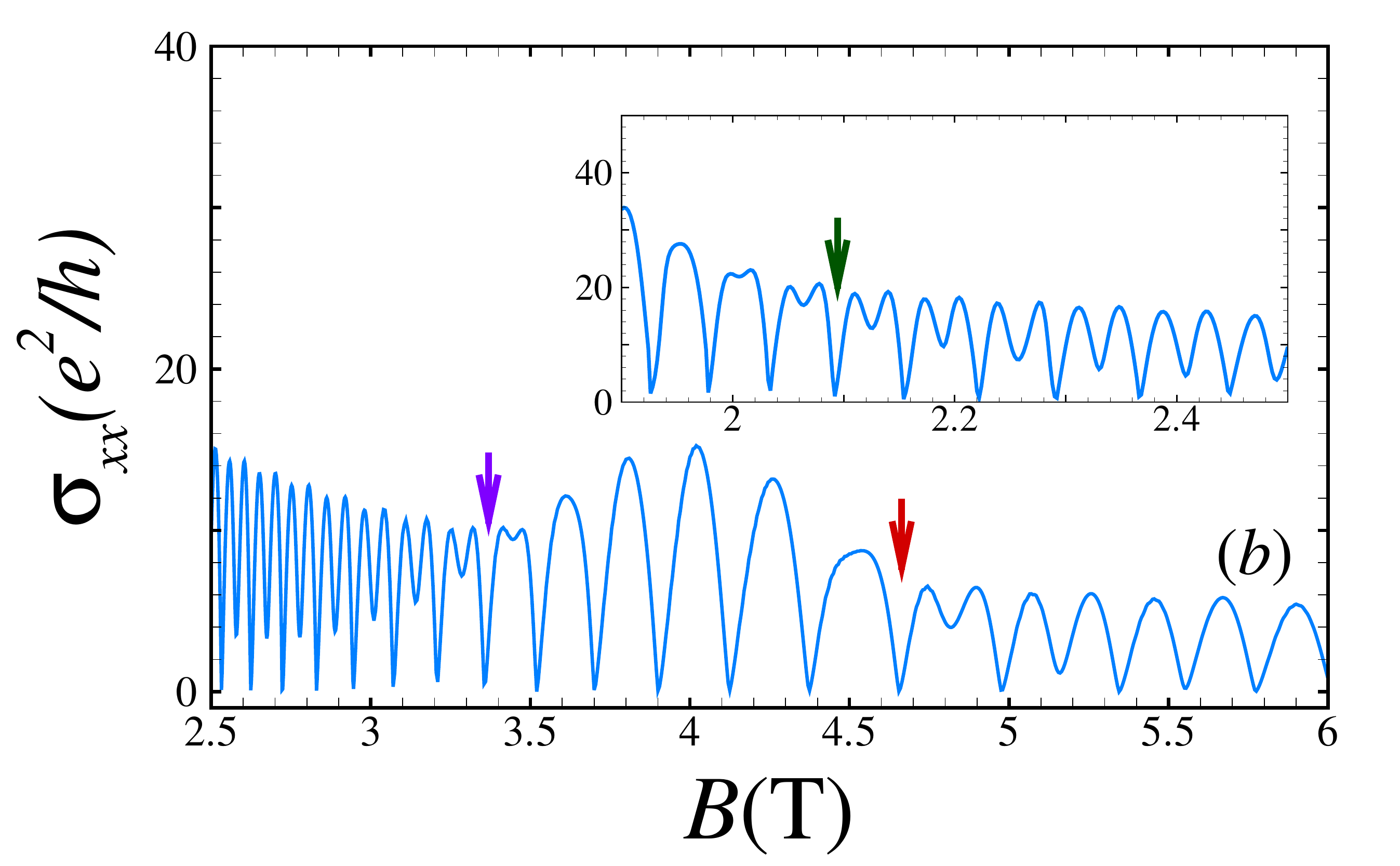}%
  }
\caption{\label{fig6} (Color online) Hall $(a)$ and longitudinal $(b)$ conductivities as functions of magnetic field at $T=50\, \rm{mK}$. The insets are the same as figures but for lower magnetic field. The arrows show the onsets of the changes in the behavior of the graphs.}
\end{figure}

In order to better understand the phenomenon, we concentrate on a weak magnetic field regime in Fig. \ref{fig6}, where we plot the longitudinal conductivity in Fig. \ref{fig6}(a) as well as the Hall conductivity of the same system in Fig. \ref{fig6}(b) with insets in both figures illustrating the lower magnetic field calculations. The interesting point about the Hall conductivity
plateaus (plotted in units of $e^2/h$) is that decreasing the magnetic field and moving from spin-split peaks to the beating regime in longitudinal conductivity figure at $B\approx 4.7\,\rm{T}$, the Hall plateaus also change in a way that
the integer Hall plateaus (with both even and odd filling factors) are replaced by the odd ones. The reverse phenomenon can be traced at $B\approx 3.4\,\rm{T}$ and once more at $B\approx2.1\,\rm{T}$ the same process begins. When the Landau levels of
two branches of the system are far enough from each other, both odd and even plateaus can be distinguished in the system, but as the magnetic field is decreased, the diverse growing rate of Landau level energies with respect to the
feeling factor leads to the condition where two subband energies approach each other and in a region where they are so close, we will have spin degenerate levels with only odd filling factors. For example at $B\approx 2.5\,\rm{T}$
where we expect an even Hall plateau,  the energy difference between adjacent
Landau levels is about $0.014\,\rm{meV}$ with Landau level broadening being $\Gamma\approx 0.021\,\rm{meV}$. The interplay between large broadening and small level splitting results in the disappearance of the Hall plateau. As we lower the magnetic field the same approach is repeated. Note that for the spin-split
peaks to appear in SdH oscillations (as well as even-odd filling factors), we need a small Landau level broadening and low-temperature condition.

\begin{figure}[h]
\centering
      \includegraphics[width=1.\linewidth] {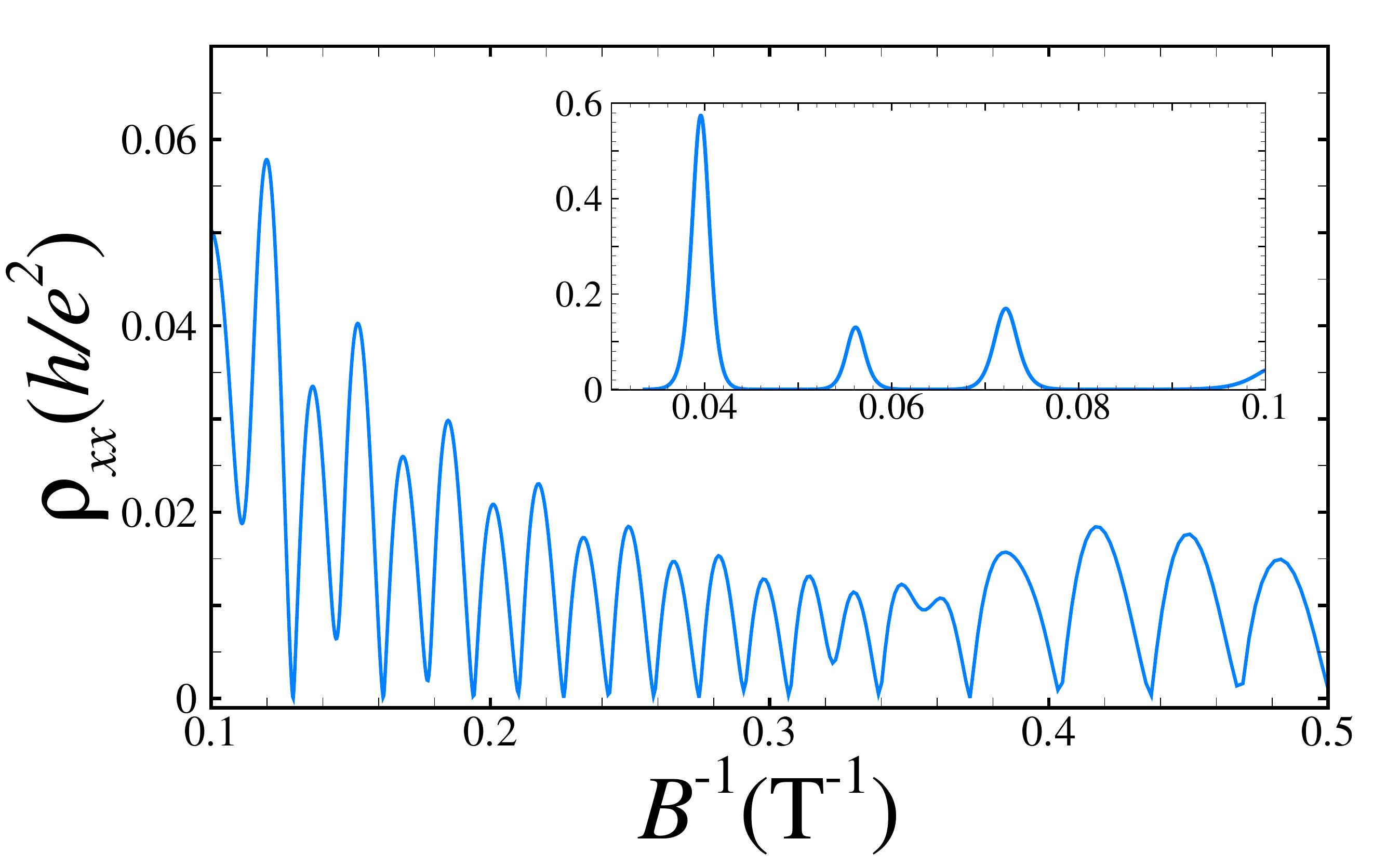}
\caption{\label{fig7} (Color online)  Magnetoresistivity as a function of $B^{-1}$ for $n_c=1.5\times10^{16}\rm{m^{-2}}$, $\alpha_0=11.1\,\rm{eV\AA^3}$ and $T=50\, \rm{mK}$. Inset: The same as in figure but for larger magnetic fields}
\end{figure}

To compare our results with SdH experiments, we plot the magnetoresistivity of 
the system versus $1/B$ in Fig. \ref{fig7}. The figure is plotted for 
$T=100\rm{mK}$, $\Gamma=0.01\sqrt{B}\,\rm{meV}$, $\alpha_0=11.1\,\rm{eV\AA^3}$ 
and $n_c=\rm{1.5\times 10^{16}~m^{-2}}$ in accordance with the densities 
reported in experiments. The inset figure also shows the high field 
magnetoresistivity. The oscillation pattern in this figure is qualitatively 
consistent with that observed in the experiments especially at larger magnetic 
fields~\cite{xie,fete,yang}. We should note that the smallest magnetic field for 
which the SdH oscillations have been observed in the present system is about 
$2\,\rm{T}$. As we stated before for low density regime the beatings occur in 
lower magnetic fields, which is due to a small Landau level separation in the 
present system, it is not easy to be detected in experiments.

\begin{figure}[h]
\centering
\subfloat{%
  \includegraphics[width=1.\linewidth]{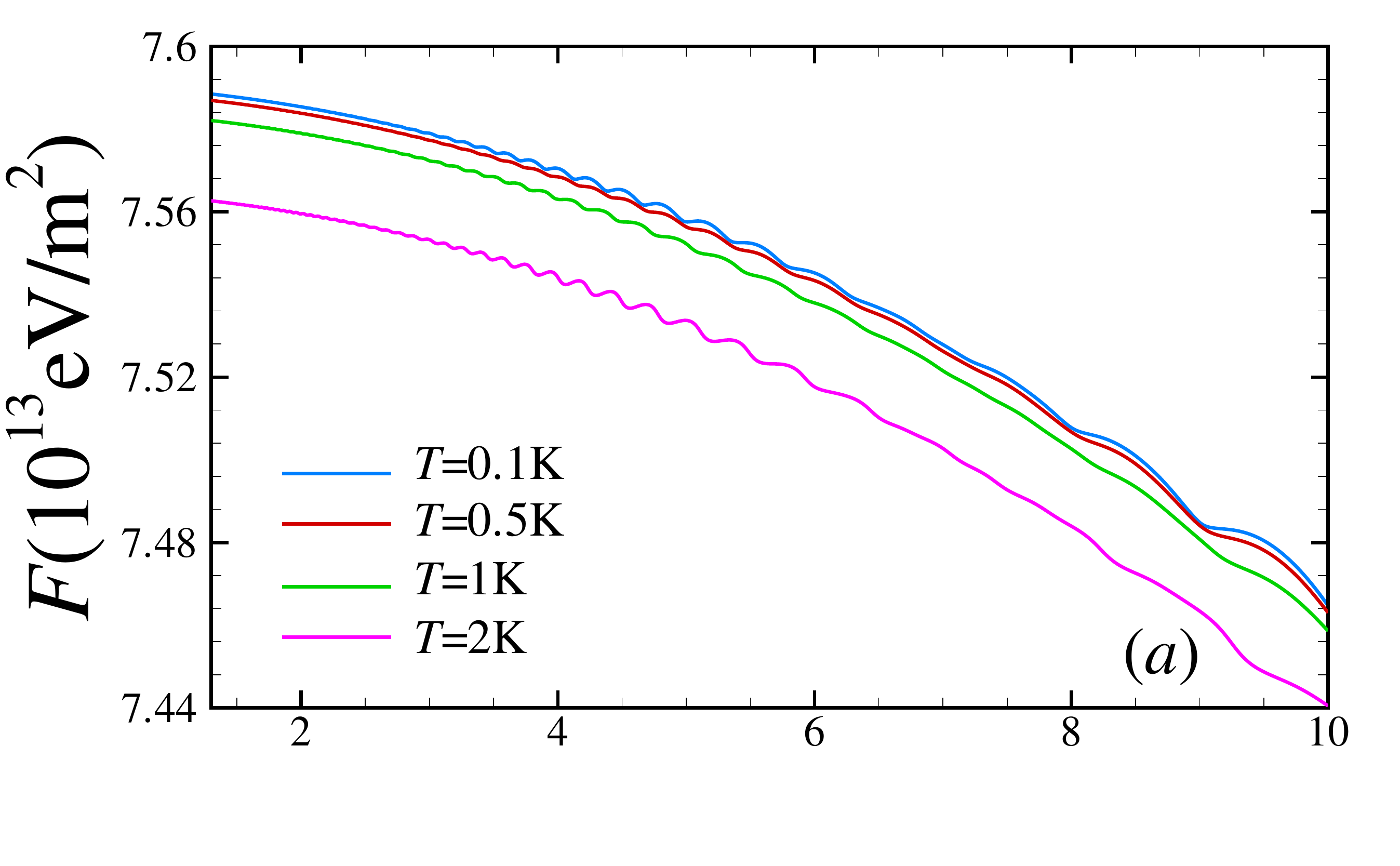}%
  }\par 
\negthickspace\negthickspace\negthickspace\negthickspace\negthickspace\negthickspace\negthickspace\negthickspace\negthickspace\negthickspace\negthickspace\negthickspace\negthickspace
       
\subfloat{%
  \includegraphics[width=1.\linewidth]{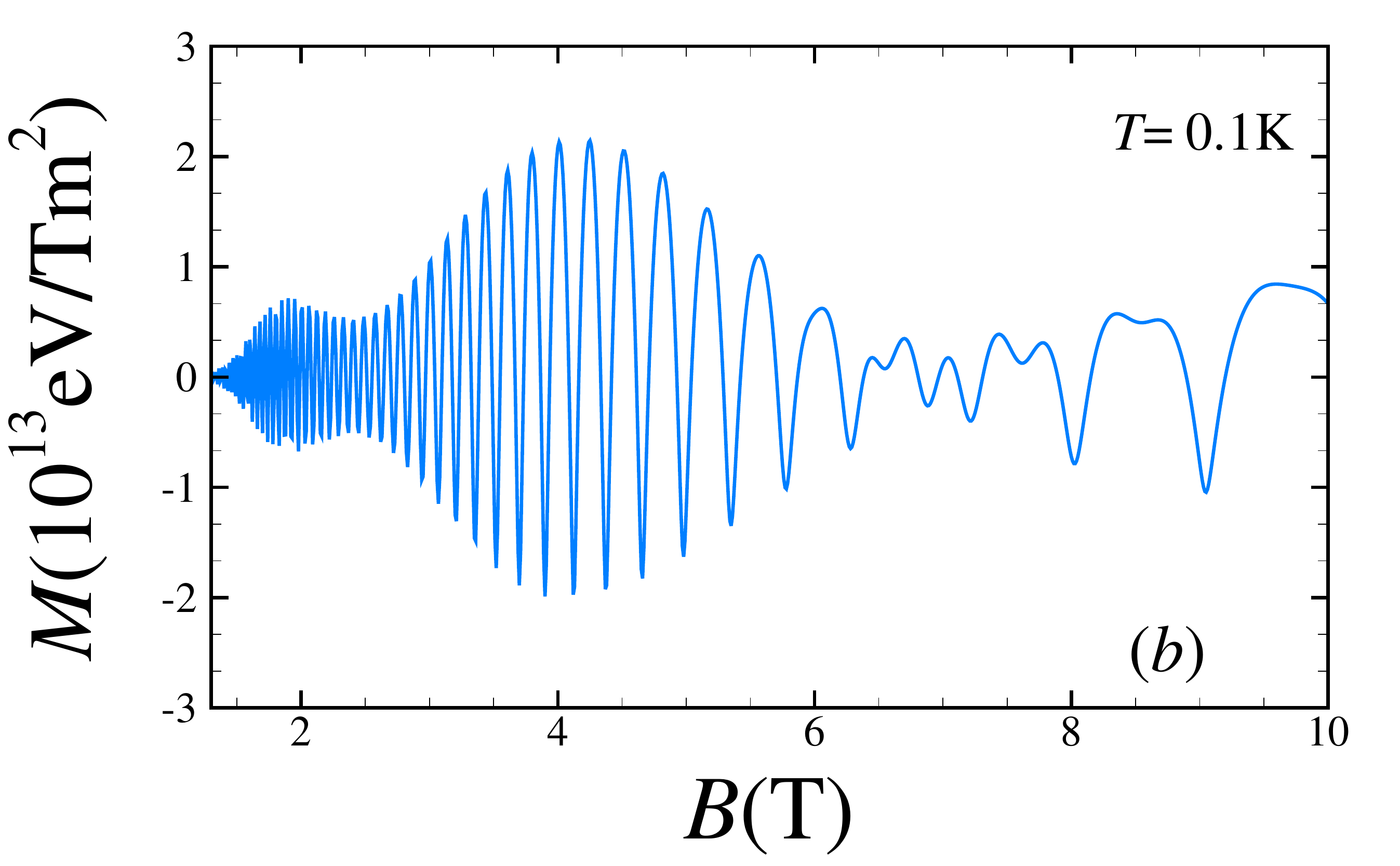}%
  }
\caption{\label{fig8} (Color online)  (a) Free energy of the system $F$ versus the magnetic field $B$ for different temperatures. (b) Magnetization oscillations $M$ versus magnetic field $B$ at $T=100\, \rm{mK}$. We consider $n_c=3.5\times10^{16}\rm{m^{-2}}$, $\alpha_0=5.5\,\rm{eV\AA^3}$ and $\Gamma=0.03\sqrt{B}\, \rm{meV}$ .}
\end{figure}

It is also interesting to trace the magneto-oscillations of the system in the 
analogous de Haas-van Alphen oscillations, which are the oscillations of the 
magnetization of the electron gas defined as
\begin{equation}
M=-\frac{\partial F}{\partial B}|_{n_c,T},
\end{equation}
where
\begin{equation}
\begin{split}
F(T)=&\int d\varepsilon D(\varepsilon)\lbrace \varepsilon f(\varepsilon)\\
&+k_BT[f(\varepsilon)\rm{ln}(f(\varepsilon))+(1-f(\varepsilon))\rm{ln}
(1-f(\varepsilon))]\rbrace,
\end{split}
\end{equation}
is the free energy per unit area at temperature $T$. In Fig. \ref{fig8}(a) we illustrate the free energy of the system as a function of magnetic field for different temperatures. The oscillations of the free energy also follow the oscillations in the chemical potential, although it is hard to be resolved especially for higher temperatures. These oscillations are more clear in magnetization shown in Fig. \ref{fig8}(b). The free energy exhibits kinks and the
magnetization has jumps at those values of the magnetic field, where a Landau 
level becomes
completely empty. The parameters in this figure are the same as in Fig. \ref{fig3}(a) and we can see the same pattern in the magnetization oscillations as well.

\section{Concluding Remarks}

We have numerically evaluated the magnetotransport coefficients of the 2DEG at the interface of $\rm{LaAlO_3}$ and $\rm{SrTiO_3}$. A perturbative approach is also suggested, which could be used for lower densities and not very small magnetic fields. We have also discussed the role of the anisotropic k-cubic
Rashba spin splitting as well as the Zeeman splitting. A beating pattern is found in low magnetic field regime caused by the presence of Rashba term in the
Hamiltonian. The somehow anomalous nature of beatings in comparison with beatings caused by isotropic k-cubic Rashba interaction is illustrated. The
impact of physical characteristics of the system such as the Landau level
broadening $\Gamma$, Rashba constant $\alpha_0$ and electron concentration $n_c$
is considered as well. We have found that the beating pattern persists up to larger magnetic fields for higher densities and also stronger Rashba interactions. The Hall plateaus of the system are also plotted and the expected coincidence of the steps between plateaus in the Hall resistivity with the peaks of the longitudinal resistivity is shown.

A quantitative agreement between the SdH oscillations in our model with the experimental reports is found for somehow larger magnetic fields and low
electron densities. We would like to emphasize that the hallmark of our model
is the appearance of a peculiar beating pattern in low magnetic fields, which can not be detected in low-density regime, according to our calculations. On
the other hand for quantum oscillations to be observed, the condition
$\hbar\omega_c>k_BT$ should be fulfilled and due to the larger effective mass in
the present system in comparison with conventional 2DEGs, lower temperatures for
detecting the SdH oscillations and Hall plateaus in the typical ranges of magnetic fields is essential.

--

\section{Acknowledgements}
R. A would like to thank the international center of theoretical physics, ICTP, Italy where the final stage of this work has been completed. This work is supported by the Iran Science Elites Federation grant no 11/66332.

\appendix

\section{}\label{ap-one}
Using the approximate eigenvalues and eigenfunctions \eqref{eq:eig1}-\eqref{eq:psi_-} we can obtain the following expressions for $|F_{n}^{+}(u)|^2$ and $|F_{n}^{-}(u)|^2$

\begin{equation}\label{eq:F_+}
\begin{split}
&|F_n^+(u)|^2=|\bra{\psi_n^+}e^{i\mathbf{q}.\mathbf{r}}\ket{\psi_n^+}|^2\\
&=e^{-u}N_{n,+}^4[L_n(u)+C_{n,+}^2 L_{n-1}(u)+D_{n,+}^2 L_{n+3}(u)\\
&+2\sqrt{{(n-1)!}/{(n+3)!}}C_{n,+}D_{n,+}u^2 L_{n-1}^{(4)}(u)]^2 \delta_{k_y,k'_y+q_y},
\end{split}
\end{equation}
\begin{equation}\label{eq:F_-}
\begin{split}
&|F_n^-(u)|^2=|\bra{\psi_n^-}e^{i\mathbf{q}.\mathbf{r}}\ket{\psi_n^-}|^2\\
&=e^{-u}N_{n,-}^4[L_n(u)+C_{n,-}^2 L_{n+1}(u)+D_{n,-}^2 L_{n-3}(u)\\
&+2\sqrt{{(n-3)!}/{(n+1)!}}C_{n,-}D_{n,-}u^2 L_{n-3}^{(4)}(u)]^2 \delta_{k_y,k'_y+q_y},
\end{split}
\end{equation}

Here $L_n^{(m)}$ is the generalized Laguerre polynomial and as defined before $u=q^2l_c^2/2$. After a lengthy calculation we find the following expressions for $I_n^+$ and $I_n^-$

\begin{equation}
\begin{split}
&\frac{I_n^+}{N_{n,+}^4}=(2n+1)+C_{n,+}^4(2n-1)+D_{n,+}^4(2n+7)\\
&+4C_{n,+}^2D_{n,+}^2(2n+3)
-C_{n,+}^2n+12\frac{C_{n,+}D_{n,+}}{N_{n,+}^2}\sqrt{\frac{(n-1)!}{(n+3)!}},
\end{split}
\end{equation}
\begin{equation}
\begin{split}
&\frac{I_n^-}{N_{n,-}^4}=(2n+1)+C_{n,-}^4(2n+3)+D_{n,-}^4(2n-5)\\
&+4C_{n,-}^2\!D_{n,-}^2\!(2n\!-\!1)\!
-\!C_{n,\!-}^2\!(n+\!\!1)\!+\!12\frac{C_{n,-}\!D_{n,-}}{N_{n,-}^2}\sqrt{\frac{(n\!-\!3)!}{(n\!+\!1)!}},
\end{split}
\end{equation}
\section{}\label{ap-two}
The matrix elements of the velocity operator read as
\begin{widetext}
\begin{eqnarray}
&&
\begin{split}
&\bra{\psi_n ^+} \upsilon_y\ket{\psi_{n'}^+}=-i\bra{\psi_n ^+} \upsilon_x\ket{\psi_{n'}^+}=\delta_{k_yk'_y}(N_{n,+}N_{n',+})\times\\
&\lbrace(D_{n,+}B_1)\delta_{n',n+5}+(A_1+D_{n,+}B_2)\delta_{n',n+3}+(A_2+C_{n,+}B_1+D_{n,+}B_3)\delta_{n',n+1}\\
&+(A_3+C_{n,+}B_2+D_{n,+}B_4)\delta_{n',n-1}+(A_4+C_{n,+}B_3)\delta_{n',n-3}+(A_5+C_{n,+}B_4)\delta_{n',n-5}\rbrace,
\end{split}
\\[10pt]
&&
\begin{split}
&\bra{\psi_n^-}\upsilon_y\ket{\psi_{n'}^+}=-i\bra{\psi_n^-}\upsilon_x\ket{\psi_{n'}^+}=\delta_{k_yk'_y}(N_{n,-}N_{n',+})\times\\
&\lbrace(C_{n,-}A_1)\delta_{n',n+4}+(C_{n,-}A_2+B_1)\delta_{n',n+2}
+(D_{n,-}A_1+C_{n,-}A_3+B_2)\delta_{n',n}\\
&+(D_{n,-}A_2+C_{n,-}A_4+B_3)\delta_{n',n-2}+(D_{n,-}A_3+C_{n,-}A_5+B_4)\delta_{n',n-4}+(D_{n,-}A_4)\delta_{n',n-6}+(D_{n,-}A_5)\delta_{n',n-8}\rbrace,
\end{split}
\end{eqnarray}
\end{widetext}
where the constants $A_i$ and $B_i$ have different values for $\bra{\psi_n ^+} \upsilon_y\ket{\psi_{n'}^+}$ and $\bra{\psi_n ^+} \upsilon_x\ket{\psi_{n'}^+}$  such that $A_1=3\lambda C_{n',+}\sqrt{n'-1}\sqrt{n'-2}$, $A_2=\pm\gamma \sqrt{n'}\pm\lambda C_{n',+}(2n'-1)$, $A_3=\gamma\sqrt{n'+1}+\lambda C_{n',+}\sqrt{n'}\sqrt{n'+1}+3\lambda D_{n',+}\sqrt{n'+3}\sqrt{n'+2}$, $A_4=\pm\lambda D_{n',+}(2n'+7)$, $A_5=\lambda D_{n',+}\sqrt{n'+5}\sqrt{n'+4}$, $B_1=-\lambda\sqrt{n'}\sqrt{n'-1}\pm\gamma C_{n',+}\sqrt{n'-1}$, $B_2=\mp\lambda(2n'+1)+\gamma C_{n',+}\sqrt{n'}$, $B_3=-3\lambda\sqrt{n'+2}\sqrt{n'+1}\pm\gamma D_{n',+}\sqrt{n'+3}$ and $B_4=\gamma D_{n',+}\sqrt{n'+4}$. All the upper signs correspond to $\bra{\psi_n ^+} \upsilon_y\ket{\psi_{n'}^+}$ and the lower signs correspond to $\bra{\psi_n ^+} \upsilon_x\ket{\psi_{n'}^+}$.

In the same way we will have

\begin{widetext}
\begin{eqnarray}
&&
\begin{split}
&\bra{\psi_n ^-} \upsilon_y\ket{\psi_{n'}^-}=-i\bra{\psi_n ^-} \upsilon_x\ket{\psi_{n'}^-}=\delta_{k_yk'_y}(N_{n,-}N_{n',-})\times\\
&\lbrace(C_{n,-}A_1+B_1)\delta_{n',n+5}+(C_{n,-}A_2+B_2)\delta_{n',n+3}+(D_{n,-}A_1+C_{n,-}A_3+B_3)\delta_{n',n+1}\\
&+(D_{n,-}A_2+C_{n,-}A_4+B_4)\delta_{n',n-1}+(D_{n,-}A_3+B_5)\delta_{n',n-3}+(D_{n,-}A_4)\delta_{n',n-5}\rbrace,
\end{split}
\\[10pt]
&&
\begin{split}
&\bra{\psi_n ^+} \upsilon_y\ket{\psi_{n'}^-}=-i\bra{\psi_n ^+} \upsilon_x\ket{\psi_{n'}^-}=\delta_{k_yk'_y}(N_{n,+}N_{n',-})\times\\
&\lbrace(D_{n,+}B_1)\delta_{n',n+8}+(D_{n,+}B_2)\delta_{n',n+6}+(A_1+C_{n,+}B_1+D_{n,+}B_3)\delta_{n',n+4}\\
&+(A_2+C_{n,+}B_2+D_{n,+}B_4)\delta_{n',n+2}
+(A_3+C_{n,+}B_3+D_{n,+}B_5)\delta_{n',n}+(A_4+C_{n,+}B_4)\delta_{n',n-2}+(C_{n,+}B_5)\delta_{n',n-4}\rbrace,
\end{split}
\end{eqnarray}
\end{widetext}
With $A_1=\pm\gamma D_{n',-}\sqrt{n'-3}$, $A_2=\gamma D_{n',-}\sqrt{n'-2}+3\lambda\sqrt{n'}\sqrt{n'-1}$, $A_3\!=\pm\gamma C_{n',-}\sqrt{n'+1}\pm\lambda(2n'+1)$, $A_4=\gamma C_{n',-}\sqrt{n'+2}+\lambda\sqrt{n'+2}\sqrt{n'+1}$, $B_1=\lambda D_{n',-}\sqrt{n'-3}\sqrt{n'-4}$, $B_2=\mp\lambda D_{n',-}(2n'-5)$, $B_3=-3\lambda D_{n',-}\sqrt{n'-2}\sqrt{n'-1}-\lambda C_{n',-}\sqrt{n'+1}\sqrt{n'}\pm\gamma\sqrt{n'}$, $B_4=\gamma\sqrt{n'+1}\mp\lambda C_{n',-}(2n'+3)$ and $B_5=-3\lambda C_{n',-}\sqrt{n'+3}\sqrt{n'+2}$.


\begin{thebibliography}{7}
\bibitem{m-alpha}
$m^*$ and $\alpha_0$ are evaluated from their definitions in Ref.~\onlinecite{zhou1} with numerical values adapted from Ref.~\onlinecite{zhong}
\end{thebibliography}
\end{document}